\pdfoutput=1

\documentclass{article}
\usepackage{spconf}
\usepackage{amssymb,amsmath,bm}
\usepackage{multirow,makecell}
\usepackage[pdftex]{graphicx}
\usepackage[bibstyle=ieee,sorting=none,giveninits=true,citestyle=numeric-comp]{biblatex}

\def\eg{\textit{e.g.\ }}
\def\ie{\textit{i.e.\ }}

\usepackage{xspace}
\def\pa{presence\slash absence\xspace}

\title{\vspace{-1em} COMPARING THE MAX AND NOISY-OR POOLING FUNCTIONS IN MULTIPLE INSTANCE LEARNING FOR WEAKLY SUPERVISED SEQUENCE LEARNING TASKS}

\name{Yun Wang$^\dag$, Juncheng Li\,$^{\dag,\ddag}$, Florian Metze$^\dag$
\thanks{This work was supported in part by a gift award from Robert Bosch LLC
and a faculty research award from Google.
\newline \indent \hspace{0.6em}
This work used the ``bridges'' cluster (at PSC) and the ``comet'' cluster (at SDSC) of the
Extreme Science and Engineering Discovery Environment (XSEDE) \cite{XSEDE},
supported by NSF grant number ACI-1548562.}}
\address{$^\dag$Language Technologies Institute, Carnegie Mellon University, Pittsburgh, PA, U.S.A. \\
$^\ddag$Research and Technology Center, Robert Bosch LLC, Pittsburgh, PA, U.S.A. \\
{\small \tt \{yunwang, junchenl, fmetze\}@cs.cmu.edu}}

\begin{document}
\ninept

\maketitle

\begin{abstract}
Many sequence learning tasks
require the localization of certain events in sequences.
Because it can be expensive to obtain strong labeling that
specifies the starting and ending times of the events,
modern systems are often trained with weak labeling
without explicit timing information.
Multiple instance learning (MIL) is a popular framework
for learning from weak labeling.
In a common scenario of MIL, it is necessary to choose
a pooling function to aggregate the predictions for
the individual steps of the sequences.
In this paper, we compare the ``max'' and ``noisy-or'' pooling functions
on a speech recognition task and a sound event detection task.
We find that max pooling is able to localize phonemes and sound events,
while noisy-or pooling fails.
We provide a theoretical explanation of
the different behavior of the two pooling functions
on sequence learning tasks.
\end{abstract}


\section{Introduction}

Many machine learning tasks take sequences as input.
The sequences may be either text (\eg machine translation),
audio (\eg speech recognition), or in other forms.
We call such tasks \emph{sequence learning}.
Sequence learning tasks often ask for the
\emph{localization} of certain events in the sequences.
For example, in speech recognition, it is desirable
for the system to output the timespan of each word or phoneme
besides the transcription;
in sound event detection (SED), it is often required
to detect the onset and offset times of each sound event occurrence.

Systems capable of localization are conventionally trained
with \emph{strong labeling}: the annotation of the training data
specifies the timespan of each event in the sequences.
However, it is tedious to generate strong labeling for data by hand,
which is a serious limitation when scaling up the systems.
To overcome this problem, researchers have turned to
large data with \emph{weak labeling}, in which
the timespans of the events are not explicitly given.
The weakness of labels may still be classified into different levels.
For example, speech recognition systems trained with
connectionist temporal classification (CTC) \cite{CTC}
take phoneme sequences as training labels.
These labels do not specify the start and ending times
of each phoneme, but do specify the order between phonemes;
we call such labeling \emph{sequential labeling}.
Google Audio Set \cite{AudioSet}, a large corpus for SED
released in early 2017, only labels which types of sound events
are present in each recording;
we call such labeling \emph{\pa labeling},
and this is the focus of this paper.

A popular solution for learning with \pa labeling
is multiple instance learning (MIL) \cite{amores2013multiple}.
For example, MIL has been applied to SED
in \cite{su2017weakly, kumar2016audio, raj2017audio, salamon2017dcase}.
In a common scenario of MIL, it is important to choose
a pooling function to aggregate instance-level predictions
(see Fig.~\ref{fig:MIL}).
In this paper, we compare two pooling functions:
the ``max'' pooling function used in \cite{su2017weakly, kumar2016audio, salamon2017dcase},
and a ``noisy-or'' pooling function which has been
applied to object detection in images
\cite{maron1998framework, zhang2006multiple, babenko2008simultaneous}.
We evaluate the two pooling functions
on a speech recognition task and a sound event detection task,
both with \pa labeling.
We measure their performance quantitatively
as well as inspect their behavior of localization.
It turns out that max pooling learns to localize successfully,
while noisy-or pooling fails.
We provide a theoretical analysis of this result,
with special attention to the particularities of
sequence learning tasks.

\section{Multiple Instance Learning}

\begin{figure}[t]
\centering
\includegraphics[width=\linewidth]{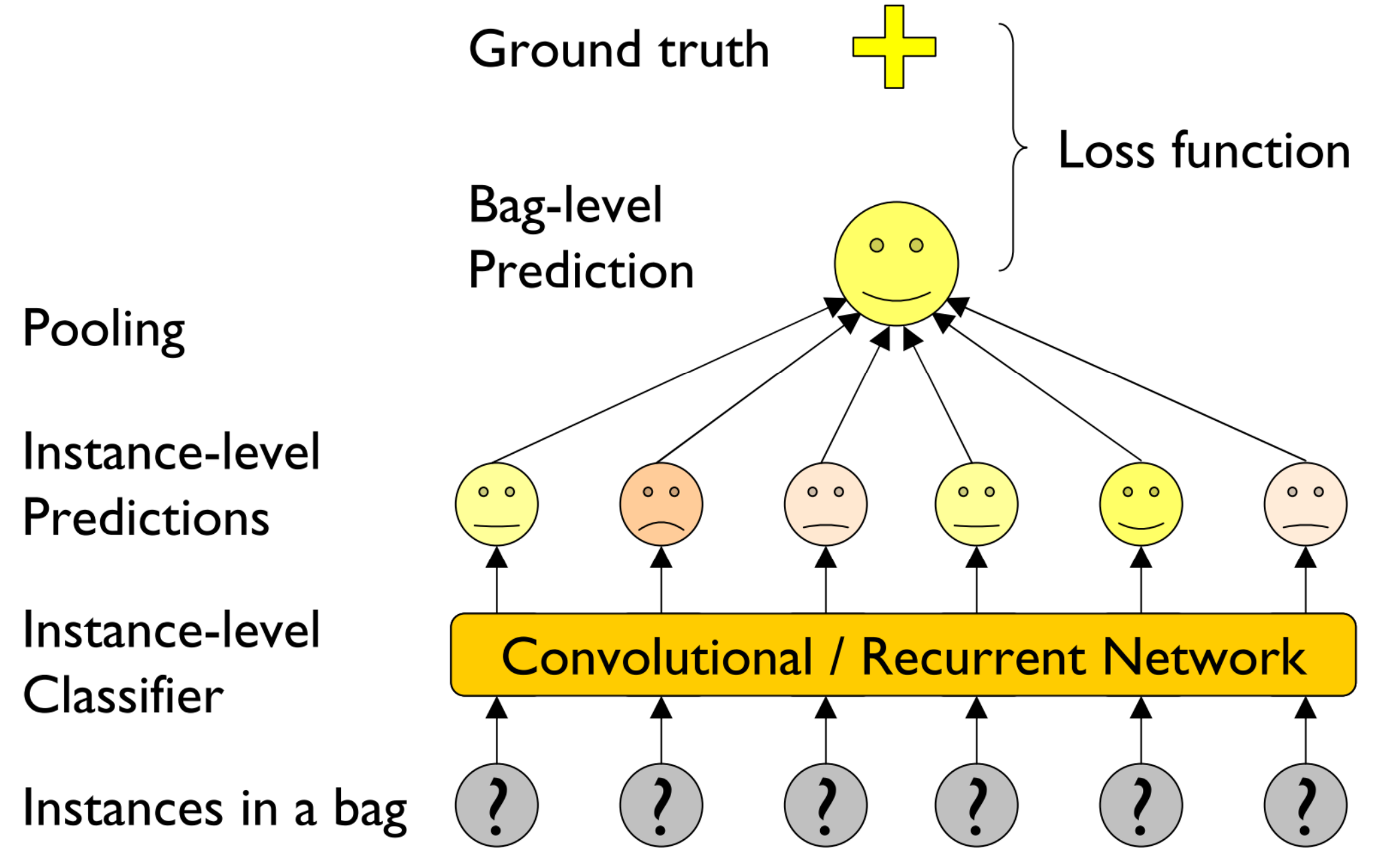}
\caption{Block diagram of a MIL system for sequence learning tasks.}
\label{fig:MIL}
\end{figure}

Many weakly supervised tasks can be formulated as
multiple instance learning (MIL) problems.
In MIL, we do not have the ground truth label for each instance;
instead, the instances are grouped into \emph{bags},
and we only have labels for the bags.
The relationship between the label of a bag and the
(unknown) labels of the instances in it often follows the
\emph{standard multiple instance (SMI) assumption}:
a bag is positive if it contains at least one positive instance,
and negative if it only contains negative instances.
For a weakly supervised sequence learning problem,
we can treat each sequence as a bag,
and the steps (\eg audio frames) of the sequence as instances.
For example, weakly supervised sound event detection
entails making frame-level predictions
in order to localize the sound events,
while only recording-level labels are given.

The structure of a MIL system using the instance space paradigm \cite{amores2013multiple}
is shown in Fig.~\ref{fig:MIL}.
The individual instances in a bag are first fed into an instance-level classifier.
The classification does not have to be independent among the instances;
for sequence learning, the instance-level classifier
can be replaced by a convolutional or recurrent neural network
that propagates information from one instance to the next.
The instance-level predictions are aggregated into a
bag-level prediction using a pooling function.
The loss function can be constructed by comparing the
bag-level prediction with the label of the bag (\eg cross-entropy),
and minimized with any optimization algorithm (\eg stochastic gradient descent).

A number of pooling functions have been proposed.
In this paper, we specifically study the ``max'' \cite{su2017weakly, kumar2016audio, salamon2017dcase}
and ``noisy-or'' \cite{maron1998framework, zhang2006multiple, babenko2008simultaneous} pooling functions.
Let $y_i \in [0,1]$ be the prediction for the \mbox{$i$-th} instance in a bag,
and $y \in [0,1]$ be the bag-level prediction.
The max pooling function simply takes the maximum instance-level
prediction as the bag-level prediction:
\begin{equation}
y = \max_i y_i, \label{eq:max}
\end{equation}
while the noisy-or pooling function treats $y_i$ as
the probability of the \mbox{$i$-th} instance being positive,
and computes the probability of the bag being positive
according to the SMI assumption:
\begin{equation}
y = 1 - \prod_i (1 - y_i). \label{eq:noisy-or}
\end{equation}

In the general MIL framework, the noisy-or pooling function
appears more principled.
Assuming that instances within a bag are mutually independent,
it provides a probabilistic interpretation of the predictions.
Also, from the perspective of error back-propagation,
all the instances in a bag can receive an error signal
(as opposed to only the maximum instance when using the
max pooling function).
However, in the situation of sequence learning,
these merits of the noisy-or pooling function may be questioned:
the steps of a sequence are often correlated with each other,
and the convolutional or recurrent neural network
functioning as the instance-level classifier
is able to propagate the gradient across time
even though only a single step receives
an error signal from higher layers.
Whether these differences are significant enough
to nullify the advantages of the noisy-or pooling function
remains to be discovered by experiment.

\section{Experiment on Speech Recognition}

State-of-the-art speech recognition systems often employ a CTC output layer \cite{CTC},
which takes phoneme sequences as training labels (sequential labeling).
We take a step further and test out the ``max'' and ``noisy-or'' pooling functions
using the even weaker \pa labeling, \ie we only tell the system
which phonemes are present in each utterance.

We conducted the experiments on the TEDLIUM v1 corpus%
\footnote{The corpus can be downloaded at \scalebox{0.95}[1.0]{\url{http://www.openslr.org/resources/7/}}.}.
The corpus consists of 206~hours of training data,
1.7~hours of development data, and 3.1~hours of testing data;
we used 95\% of the training data for training, and 5\% for validation.
We generated ground truth phoneme sequences for all utterances
from the transcriptions and the dictionary;
we only retained the 39 ``real'' phonemes and discarded all
noise markers like ``breath'' and ``cough''.

The baseline system we compared against was a Theano \cite{Theano}
re-implementation of the example CTC system%
\footnote{\scalebox{0.95}[1.0]{\url{https://github.com/srvk/eesen/tree/master/asr_egs/tedlium/v1}}}
in the EESEN toolkit \cite{EESEN}.
This system took phoneme sequences as training labels.
The network consisted of five bidirectional LSTM layers,
with 320~memory cells in each direction of each layer.
The input layer had 40~neurons, which accepted 40-dimensional filterbank features%
\footnote{Unlike the EESEN system, we did not use delta and double delta features.}.
The CTC output layer consisted of 40~neurons arranged in a softmax group,
corresponding to the 39~phonemes plus a blank token.

We adapted the baseline system to build two systems with \pa labeling,
using the ``max'' and ``noisy-or'' pooling functions, respectively.
The input and hidden layers were identical to the baseline system.
The output layer consisted of 39~neurons,
without the one neuron dedicated to the blank token.
These neurons used the sigmoid activation function.
Even though phonemes cannot overlap in time,
we did not want to enforce this restriction in the network.
The frame-level predictions are then aggregated across time
into an utterance-level prediction (a 39-dimensional vector).

All the three systems were trained to minimize the cross-entropy loss function,
averaged properly across different units (see Table~\ref{table:tedlium-details}).
The optimization algorithm was stochastic gradient descent (SGD)
with a Nesterov momentum of 0.9 \cite{Nesterov}.
Each minibatch contained 20,000 frames;
an epoch consisted of about 2,000 minibatches.
All the systems were trained for 24~epochs,
with the learning rate staying constant in the first 12~epochs,
and then halved in each of the next 12~epochs.
We found it essential to apply proper gradient clipping
and a large initial learning rate,
in order to ensure that the network could get through the
initial stage of instability safely and
progress with large enough steps after that.
The gradient clipping limit and initial learning rate
of each system are also listed in Table~\ref{table:tedlium-details}.

We decoded all the systems using simple \emph{best path decoding}:
choosing the most probable token (phoneme or blank) at each frame,
collapsing consecutive duplicate tokens, and removing blanks.
Since the output layer of the two weakly supervised systems
did not have a neuron for the blank token,
we set the prediction to blank if the probability of the
most probable phoneme was smaller than $0.5$.
To avoid the extra complexity introduced by the lexicon and the language model,
we evaluated all the systems using phone error rate (PER),
as listed in Table~\ref{table:tedlium-details}.

\begin{table}[t]
\centering
\setlength{\tabcolsep}{0.2em}
\resizebox{\linewidth}{!}{
\begin{tabular}{c||c||c|c||c|c|c|c}
\hline
\multirowcell{2}{\bf{System}} & \bf{Loss} & \multicolumn{2}{c||}{\bf{Hyperparameters}} & \multicolumn{4}{c}{\bf{PER}} \\
\cline{3-8}
& \bf{averaged over} & \bf{Grad.clip} & \bf{Init.LR} & \bf{Train} & \bf{Valid.} & \bf{Dev.} & \bf{Test} \\
\hline
CTC (baseline)   & Frames           & $10^{-4}$ & $3$    & $ 4.8$ & $15.4$ & $13.9$ & $14.9$ \\
Max pooling      & Utts., phonemes  & $0.01$    & $0.3$  & $40.5$ & $43.0$ & $39.7$ & $40.7$ \\
Noisy-or pooling & Frames, phonemes & $10^{-8}$ & $3000$ & $91.0$ & $91.6$ & $91.6$ & $91.5$ \\
\hline
\end{tabular}
}
\caption{The details and performance of the CTC and weakly supervised speech recognition systems.}
\label{table:tedlium-details}
\end{table}

\begin{figure}[t]
\centering
\includegraphics[width=\linewidth]{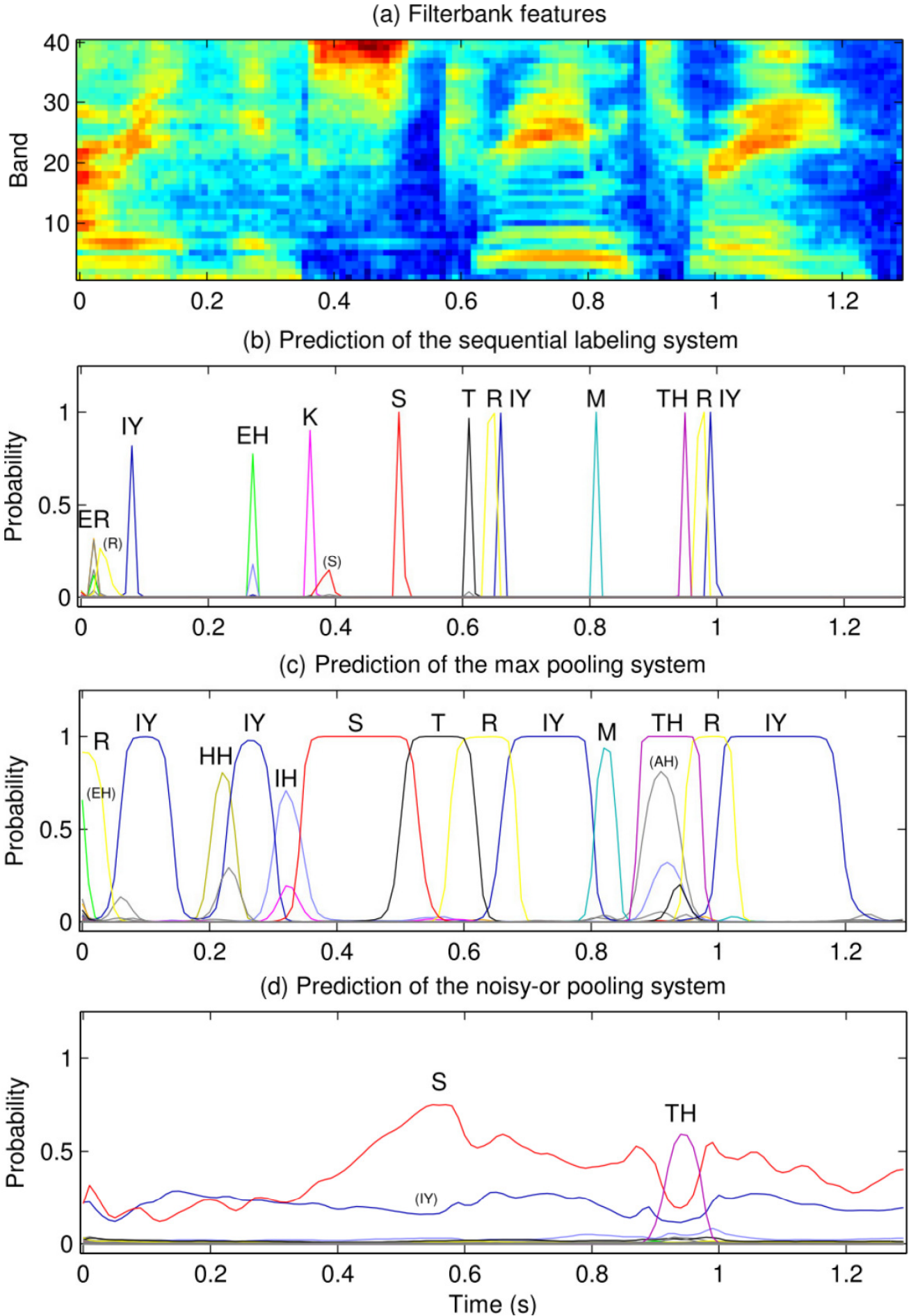}
\caption{The frame-level predictions of the various systems on an example utterance.
The ground truth transcription is ``very extreme terrain''
(both the baseline and the max pooling systems mis-recognize ``terrain'' as ``three'').
Phonemes are differentiated by color.
Peaks are annotated with their corresponding phonemes;
phonemes in parentheses were not selected in the best path decoding.}
\label{fig:tedlium-prediction}
\end{figure}

The baseline CTC system learnt fast and accurately,
reaching a PER of 30\% after the first epoch and converging to 15\%.
The max pooling system reached a PER of 43\%.
Even though there was a gap between the max pooling system and the baseline,
the learning can be regarded as successful considering that
the max pooling system only saw \pa labeling during training.
The noisy-or pooling system, however, exhibited a PER above 90\%
even after 24~epochs, and its predictions were mostly blank.

Fig.~\ref{fig:tedlium-prediction} shows the predictions of the three systems
on an example utterance.
The CTC system produces narrow peaks that align well
with the actual position of the phonemes,
but the peaks do not indicate the timespan of each phoneme.
The max pooling system produces wide peaks that clearly
indicate the onset and offset time of each phoneme,
which is exactly the desired behavior for localization.
The noisy-or pooling system fails to predict anything meaningful,
and only three phonemes receive non-negligible probabilities.

The results clearly indicate the ability of max pooling
to perform localization.
But why did noisy-or pooling fail,
despite its probabilistic interpretation and
easy propagation of error signal?
Looking into the interaction between the loss and the pooling function,
we found the noisy-or pooling system was excessively
\emph{harsh on false alarms} and \emph{lenient on misses}.

Noisy-or pooling is harsh on false alarms precisely because
it erroneously assumes that frames in a recording are independent.
Because consecutive frames in a sequence are often correlated,
when the system makes a false alarm,
it normally generates a peak that spans several frames.
This peak should be penalized only once, not for every frame it spans.
The noisy-or pooling function, however, multiplies the $1-y_i$ of each frame,
which makes the bag-level prediction $y$ extremely close to $1$,
and results in a large loss $-\log(1-y)$.
For example, in Fig.~\ref{fig:tedlium-prediction} (c),
the max pooling system makes a false alarm of the phoneme \texttt{TH}.
The frame-level prediction $y_i$ exceeds $0.999$ for 7~frames;
the maximum value is $1 - 2 \times 10^{-7}$.
With the max pooling function, this incurs a loss of
$-\log(2 \times 10^{-7}) \approx 15.5$.
If we used noisy-or pooling instead,
this false alarm would incur a loss of at least
$-\log(1 - 0.999) \times 7 \approx 48$.
As a result, the noisy-or pooling system only dared to
generate a small peak for the phoneme \texttt{TH},
as shown in Fig.~\ref{fig:tedlium-prediction} (d).

Noisy-or pooling is lenient on misses
because the system may believe it has made the correct
bag-level prediction for a positive bag,
while all the instances-level predictions are negative.
This also stems from the multiplication in the noisy-or pooling function.
Let's look at the phoneme \texttt{IH}.
Although hardly visible in Fig.~\ref{fig:tedlium-prediction} (d),
its predicted probability is around $0.02$ throughout the 130-frame utterance.
The bag-level prediction calculated by the noisy-or pooling function is
$y \approx 1 - (1 - 0.02)^{130} \approx 0.93$ --
the system believes it has predicted the presence of \texttt{IH} correctly,
and therefore will no longer make an effort to boost its frame-level probabilities.
The phoneme \texttt{IY} illustrates a more extreme case:
its predicted probability fluctuates around $0.2$,
resulting in a bag-level prediction of
$y \approx 1 - (1 - 0.2)^{130} \approx 1 - 2.5 \times 10^{-13}$.
With the bag-level prediction so close to $1$,
virtually no error signal will be passed down the network.
This problem is inherent with the noisy-or pooling function,
and limits its use to small bags.
Sequence learning tasks, however, often feature large bags
with hundreds of instances.

Max pooling, on the other hand, does not suffer from these fatal
defects of noisy-or pooling.
With an underlying RNN that can propagate information across time,
it learns to perform sequence learning tasks relatively easily.

\section{Experiments on Sound Event Detection}

We conducted experiments on Task 4 of the DCASE 2017 challenge \cite{DCASE2017}.
This task consists of two subtasks:
Task A is audio tagging, \ie determining which events are present in each recording,
but without localizing the events;
Task B is sound event detection, \ie Task A plus localization.
Both tasks consider 17~events related to vehicles and warning sounds.

The data consists of a training set (51,172~recordings),
a public test set (488~recordings),
and a private evaluation set (1,103~recordings).
All the recordings come from Google Audio Set \cite{AudioSet},
and are 10-second excerpts from YouTube videos.
The test and evaluation sets are strongly labeled so
they can be evaluated for both subtasks,
but the training set only comes with \pa labeling.
Also, the test and evaluation sets have balanced numbers of the events,
but the training set is unbalanced.
We set aside 1,142~recordings from the training set
to make a balanced validation set, and used the remaining
recordings for training.
We did not do anything about the class imbalance in the training data.

\begin{figure}[t]
\centering
\includegraphics[width=\linewidth]{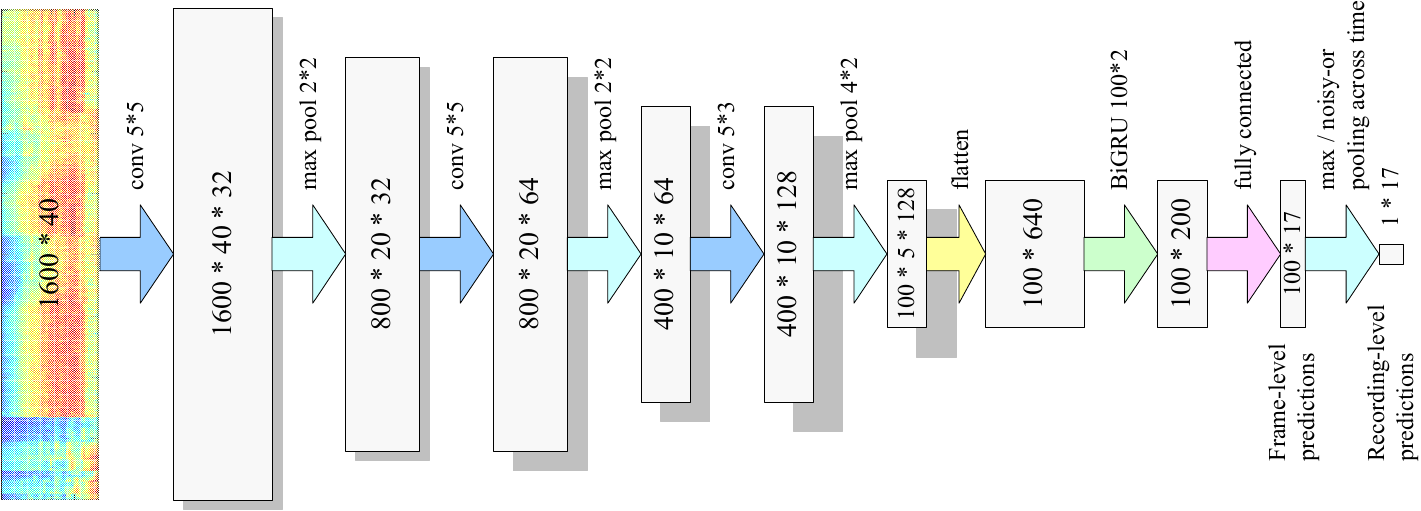}
\caption{The structure of the CRNN for audio tagging and sound event detection.
All convolutional layers use the ReLU activation.}
\label{fig:dcase-structure}
\end{figure}

We trained a convolutional and recurrent neural network (CRNN)
using the Keras toolkit \cite{Keras}.
The structure of the network is shown in Fig.~\ref{fig:dcase-structure}.
The input is 40-dimensional filterbank features of the
audio recording sampled at 160~frames per second,
\ie a matrix of size $1600 \times 40$.
The convolutional and pooling layers reduce the frame rate to 10~Hz,
whose output is fed into a bidirectional
GRU layer with 100~neurons in each direction.
A fully connected layer with 17~neurons and
the sigmoid activation function then predicts
the probability of each sound event at each frame.
These frame-level predictions can be used for SED;
they are also aggregated with either the max or the noisy-or
pooling function to produce recording-level predictions for audio tagging.
During training, the recording-level predictions
are compared against the \pa labeling of the training data
to compute the loss function.

We minimized the cross entropy averaged over both recordings and events
using the SGD algorithm with a batch size of 100~recordings.
The initial learning rate was $0.1$ for the max pooling network
and $0.3$ for the noisy-or pooling network, both with a Nesterov momentum of $0.9$.
Gradient clipping with a limit of $10^{-4}$ was found to be necessary
for noisy-or pooling, but no gradient clipping was needed for max pooling.
For the max pooling network, we applied dropout with a rate of $0.1$,
and decayed the learning rate with a factor of $0.8$
when the validation loss did not reduce for 3~consecutive epochs,
which contributed marginally to the performance.

The frame-level and recording-level probabilities predicted
by the network must be thresholded to generate output for evaluation.
Audio tagging was evaluated with the $F_1$ metric,
while SED was evaluated with both the error rate (ER) and the $F_1$ based on 1-second segments.
All the metrics were micro-averaged over all the 17~event classes.
We found it critical to tune the threshold for each event class individually.
We devised an iterative procedure to tune the class-specific thresholds
to optimize the micro-average $F_1$:
first, we tuned the threshold of each class to maximize the class-wise $F_1$;
then, we repeatedly picked a random class and re-tuned its threshold
to optimize the micro-average $F_1$, until no improvements could be made.
After each epoch of training, we tuned the thresholds on the validation data
to optimize the audio tagging $F_1$;
the model with the highest $F_1$ was picked as the final model.
The thresholds obtained on the validation data were
applied to the test and evaluation data for both audio tagging and SED.

\begin{table}[t]
\centering
\begin{tabular}{c|c|c|c|c}
\hline
\bf{Task} & \bf{Metric} & \bf{Valid.} & \bf{Test} & \bf{Eval.} \\
\hline
\bf{Task A (tagging)} & \bf{F1} & $53.3$ & $50.1$ & $50.7$ \\
\hline
\multirowcell{2}{\bf{Task B (SED)}} & \bf{Seg. ER} & & $79.7$ & $75$ \\
\cline{2-5}
& \bf{Seg. F1} & & $39.4$ & $46.9$ \\
\hline
\end{tabular}
\caption{The performance of the max pooling system on both the audio tagging
and the SED subtasks of the DCASE 2017 challenge.}
\label{table:dcase-performance}
\end{table}

The performance of the max pooling system is shown in Table~\ref{table:dcase-performance}.
This is comparable with most of the participates of the challenge in both subtasks.
The noisy-or pooling system achieved a validation $F_1$ of 53.4\%
and a test $F_1$ of 49.6\% for audio tagging.
As first sight, this seems to indicate that noisy-or pooling
performs as well as max pooling.
However, we will show that this is only the case for audio tagging,
and that noisy-or pooling is not suitable for SED.

\begin{figure}[t]
\centering
\includegraphics[width=\linewidth]{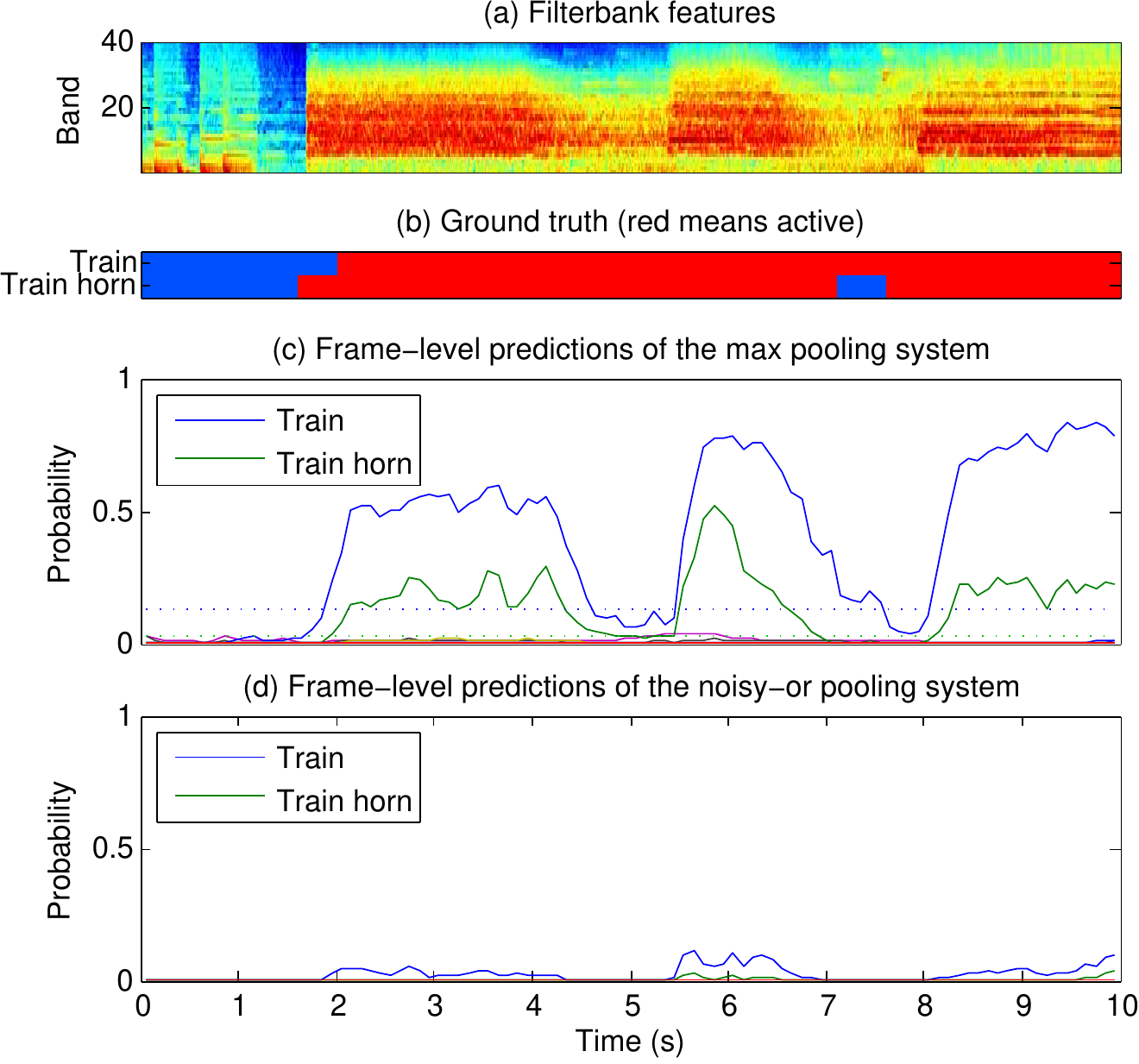}
\caption{The frame-level predictions of the max and noisy-or pooling systems
on the example test recording ``\texttt{10i60V1RZkQ}''.
In (c), the dotted lines indicate the thresholds for the two event classes.}
\label{fig:dcase-example}
\end{figure}

Fig.~\ref{fig:dcase-example} illustrates the frame-level predictions
of the max pooling and the noisy-or pooling systems on an example test recording.
The recording contains the sound of a train sounding its horn intermittently.
Fig.~\ref{fig:dcase-example}(c) indicates that the max pooling system
is able to locate the intervals during which the horn is sounding
and produce reasonable frame-level probabilities.
Fig.~\ref{fig:dcase-example}(d), however, indicates that
the noisy-or pooling system suffers from the same problem
observed in the speech recognition experiment:
its frame-level predictions are too small,
and will almost certainly be rejected even by class-specific thresholds.
Even though these small frame-level predictions can make up
reasonable recording-level probabilities through the multiplication
in the noisy-or pooling function (Eq.~\ref{eq:noisy-or}),
they are not intuitive as frame-level probabilities.

If we look closer at the frame-level predictions in Fig.~\ref{fig:dcase-example}(d),
we can notice that they seem to follow the same trend
as in Fig.~\ref{fig:dcase-example}(c),
despite their small magnitude.
Would it be possible to perform SED with these small
frame-level predictions, if we were able to set commensurate thresholds?
We used the same iterative procedure to optimize
class-specific thresholds on the test set,
and obtained a segment-based error rate of 83.5\% and an $F_1$ of 40.9\%.
Note that these are the \emph{oracle} performance of the
noisy-or pooling system, yet they are barely comparable with the
\emph{actual} performance of the max pooling system
(79.7\% ER, 39.4\% $F_1$).
For reference, the oracle performance of the max pooling system
is 75.7\% ER and 45.8\% $F_1$.

The analysis above indicates that the frame-level predictions of
the noisy-or pooling system not only have improper \emph{magnitudes},
but also lack the \emph{quality} for good SED performance.
The noisy-or pooling system is best treated as a black box
for audio tagging.

\section{Conclusion}

We evaluated the max and noisy-or pooling functions on two
sequence learning tasks with weak labeling:
phone recognition on the TEDLIUM corpus,
and audio tagging / sound event detection in the DCASE 2017 challenge.
We found that systems using the two pooling functions achieved
comparable performance in the audio tagging task,
which did not require localization.
However, only the max pooling system succeeded in localizing
the phonemes and sound events;
the noisy-or pooling system tended to produce excessively small frame-level probabilities
despite the elegant theoretical interpretation of the noisy-or pooling function.
We attribute the failure of the noisy-or pooling system to two factors:
(1) the correlation between consecutive frames violates
the assumption of independence, and
(2) the multiplication in the noisy-or pooling function can lead the system
to believe a sequence is positive even though all its frames are negative.
We recommend using the max pooling function for
weakly supervised sequence learning tasks that require localization.

\section{REFERENCES}
\label{sec:refs}

\printbibliography[heading=none]

\end{document}